\documentclass[%
reprint, 
superscriptaddress, 
showpacs,
amsmath, amssymb, 
aps, 
prl,
]{revtex4-1}
\usepackage{graphicx}
\usepackage{dcolumn}
\usepackage{bm}
\usepackage{graphicx,xcolor,float}
\usepackage[marginal]{footmisc}

\usepackage{hyperref}
\hypersetup{colorlinks=true, citecolor=blue, urlcolor=blue, linkcolor=blue}
\begin{document}
	\title{Proof-of-principle demonstration of temporally multiplexed quantum repeater link based on atomic ensemble}
	\author{Minjie Wang}
	\author{Haole Jiao}
	\author{Jiajin Lu}
	\author{Wenxin Fan}
	\author{Zhifang Yang}
	\author{Mengqi Xi}
	\author{Shujing Li} 
	
	\email{lishujing@sxu.edu.cn}
   	\author{Hai Wang}
	\email{wanghai@sxu.edu.cn}

	\affiliation{The State Key Laboratory of Quantum Optics and Quantum Optics Devices, Institute of Opto-Electronics, Shanxi University, Taiyuan 030006, China}%
	\affiliation{Collaborative Innovation Center of Extreme Optics,
		Shanxi University, Taiyuan 030006, China}
	\date{\today}
	\begin{abstract}
Duan-Lukin-Cirac-Zoller quantum repeater protocol provides a feasible scheme to implement long-distance quantum communication and large-scale quantum networks. The elementary link, namely the entanglement between two atomic ensembles, is a fundamental component of quantum repeater. For practical quantum repeater, it is required that the elementary link can be prepared with high yield and the spin waves stored in atoms can be efficiently converted into photons on demand. However, so far, such quantum repeater link has not been demonstrated in experiments. Here, we demonstrate a proof-of-principle multiplexed quantum repeater link by entangling two temporally multiplexed quantum memory. Compared with a single-mode link, the successful preparation rate of the multiplexed link is increased by one order of magnitude. By using the cavity-enhanced scheme, the on-demand retrieval efficiency of atomic spin waves is improved to 70\%, which is beneficial for the subsequent entanglement swapping between adjacent links. The realization of temporally multiplexed quantum repeater link with high retrieval efficiency lays a foundation for the development of practical quantum networks.
\end{abstract}
\maketitle
\section{Introduction}
Long-distance entanglement distribution is a fundamental task of large-scale quantum networks\cite{1,2,3,4}. Based on this, one can implement distributed quantum computation\cite{5}, remote quantum key distribution\cite{6}, and loophole-free Bell test\cite{7}.Long-distance entanglement distribution can be achieved by satellite-based free-space quantum communications\cite{8} or transferring quantum information in fiber-based channels.However, the entanglement distribution distance based on optical fiber is limited by the transmission loss of optical fiber.Quantum repeater scheme\cite{9} has been proposed to extend the communication distance, in which, the transmission channel is divided into a number of elementary links. Entanglement is first created between two end nodes of each link, and the distance of entanglement distribution is extended by entanglement swapping.

The Duan, Lukin, Cirac, and Zoller (DLCZ) protocol\cite{10}, which combines atomic ensemble and linear optics, has the advantage in entanglement generation and connection. The protocol has received extensive attention, and many related impressive advancements have been reported. For example, the storage lifetime of atomic spin-wave has been expanded to second order by combining magnetic-field-insensitive\cite{11,12,13}, long wavelength spin-wave\cite{12, 14} and confining the atoms in optical lattices\cite{15,16,17}.The retrieval efficiency of atomic memory is improved significantly in high optical-depth cold atomic ensemble\cite{18,19,20} or using the cavity-enhanced scheme\cite{14, 15, 21,22,23}. 

However, to ensure quantum correlation characteristics, the excitation probability of spontaneous Raman scattering (SRS) has to be controlled at a low level (about 1\%), which makes the long-distance communication rate incredibly low\cite{10, 24}. Multiplexing provides a solution to this problem\cite{25}. Spatially-multiplexed quantum memories have been realized by using many individually accessible memory cells with programmable addressing\cite{26}, by spatially-resolved single-photon detection\cite{27, 28} or by collecting the Stokes photon in multiple spatial directions simultaneously in cold atomic ensemble\cite{29, 30}.Temporally multiplexed DLCZ quantum interfaces have been also demonstrated by applying a train of write pulses to a cold atomic ensemble along different directions\cite{31, 32}, or applying a reversible gradient magnetic field to control the rephasing of the spin waves\cite{33, 34} in atomic ensemble. Temporal multiplexing is more resource efficient and scalable than spatial multiplexing due to its reuse of the same detectors\cite{35}.
\begin{figure*}[htp]
	\centering\includegraphics[width=14cm]{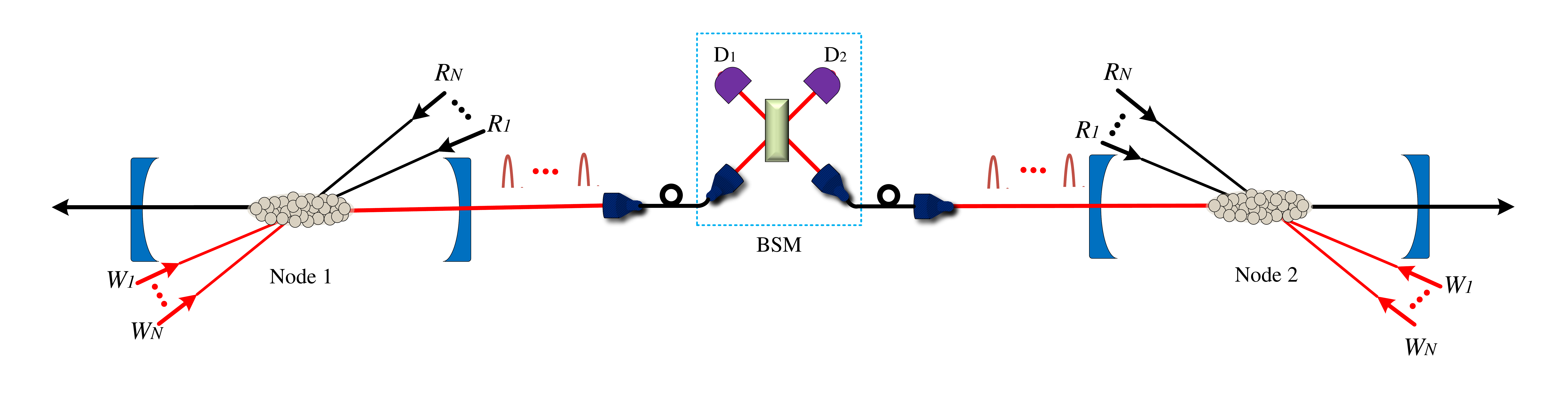}
	\caption{Quantum repeater scheme based on the cavity-enhanced and temporally multiplexed memory node 1 and node 2. A train of write pulses simultaneously act on two memory nodes from different directions, exciting Stokes photons and spin waves at both nodes with a certain probability. The emitted Stokes photons travel to the middle station of the link for bell state measurement (BSM), and successful detection of photons on D$_1$ or D$_2$ herald entanglement between the two storage nodes. W\textit{i} and R\textit{i}: the \textit{i}-th write pulse and the corresponding read pulse. D$_1$(D$_2$): single photon detector.}
\end{figure*}

At present, one of the goals of quantum repeater is how to achieve a high-quality quantum repeater link. Some pioneering progress has been made in this regard\cite{36,37,38,39,40,41,42,43,44}. Pan’s group has realized entanglement over 22 kilometers based on efficient quantum frequency conversion and cavity-enhanced emissive quantum memory\cite{39}.However, the quantum memory is single mode type, which limits the entanglement generation rate. Guo’s group demonstrated the heralded entanglement between absorptive quantum memories, in which the temporal multiplexing is utilized to enhance the entanglement distribution rate\cite{42}. Hugues de Riedmatten’s group reported telecom-heralded entanglement between two temporal multiplexing solid-state quantum memories\cite{40}.The solid-state quantum memory has a natural advantage in terms of multi-mode operation, but the spin-wave stored in atoms cannot be on demand and efficiently converted into photons the storage lifetime is short  (tens of ${\rm{\mu s}}$).

 Here, we demonstrate a temporally multiplexed elementary link, in which the spin waves stored in a cold atomic cloud can be efficiently converted into photons on demand. In the atoms, two temporally multiplexed collective excited states of atoms with different spatial modes are constructed via SRS to simulate the quantum memories of two end nodes in an elementary link. The entanglement between two quantum memories is generated via single photon interference. Compared with single mode link, the rate of entanglement generation has increased by an order of magnitude (11.8-flod). Besides, we improve the retrieval efficiency to 70\% using cavity-enhanced scheme through the Purcell effect, which can greatly improve the success probability of entanglement swapping between adjacent links. The demonstration of temporally multiplexed elementary link with on-demand and high-efficiency retrieval paves the road for practical quantum communication and quantum network.
 
\section{Cavity-enhanced and temporally multiplexed quantum repeater}

We show the temporally multiplexed quantum repeater scheme based on the cavity-enhanced quantum memory in Fig.1. The functional quantum link consists of two quantum memories, labeled node1 and node 2. In each quantum node, a $^{87}$Rb atomic ensemble is placed inside a ring cavity to carry out memory. A train of write pulses is applied on the atoms in different directions to create the correlation between atomic spin wave and Stokes photon via SRS process. The Stokes photon is resonant with cavity, and collected along cavity mode. 
To demonstrate the multiplexed quantum repeater link, the two atomic ensembles are triggered by the synchronous write pulse train. The emitted Stokes photons are sent to the middle station to perform the Bell state measurement via single photon interference. When a single photon is detected by detector D$_1$ or D$_2$, it means that entanglement between two quantum memories is successfully established. The success probability for entanglement generation in an individual elementary link with N-mode multiplexing can be evaluated by\cite{25}
\begin{equation}
	P_0^{(N)} = 1 - {(1 - {P_0})^N} \approx N{P_0}
\end{equation}
where ${P_0} = {\chi ^{}}{e^{ - {L_0}/(2{L_{att}})}}\eta _{FC}^{}\eta _{TD}^{}$ is the success probability using non-multiplexed nodes, $\chi $ is the excitation probability of Stokes photon, ${L_{att}}$ is the attenuation length of the fiber channel, ${L_0}$ is the distance between two quantum memory, $\eta _{FC}^{}$ is the memory-to-telecom frequency conversion efficiency; $\eta _{TD}^{}$ is the total detection efficiency. Only one attempt at entanglement generation can be made per communication interval ${T_{cc}} = {L_0}/c$, where c is the speed of light in fibers. So the time required to establish entanglement in a elementary link is ${t_0} \simeq {T_{cc}}/P_0^{(N)} = {T_{cc}}/\left( {N{P_0}} \right)$.

When the entanglement is generated in each quantum link, one can begin the 1-st level entanglement swapping. The atomic spin-waves are mapped into anti-Stokes photons to perform the Bell state measurement by acting a strong read pulse on the atom ensembles. To make the anti-Stokes photons propagate counterclockwise along the cavity mode, the read pulse propagates counter to the write pulse which produces the Stokes photon. The cavity can enhance the reading process through the Purcell effect and improve the retrieval efficiency significantly. The success probability for entanglement swapping at the 1-st level is ${P_1} = {R_0}{e^{ - {t_0}/{\tau _0}}}^{}\eta _{TD}^{}$, where ${\tau _0}$ is the memory lifetime in atomic ensembles, and ${R_0}$is the retrieval efficiency at zero delay. 

Followed by analogy, the success probability for entanglement swapping at the \textit{i}-th level is ${P_j} = {R_0}{e^{ - {t_{i - 1}}/{\tau _0}}}^{}\eta _{TD}^{}$, where ${t_i} \simeq {t_{i - 1}}/{P_i}$ is the time needed for the entanglement swapping at the \textit{i}-th level. It is assumed that n-level entanglement swapping is required to distribute entanglement to the whole communication channel. Then the communication rate can be expressed as:
\begin{equation}
{R_{rate}} \approx \frac{1}{{{T_{cc}}}}P_0^{(N)}\left( {\prod\nolimits_{i = 1}^{i = n} {{P_i}} } \right){P_{pr}}
\end{equation}
where ${P_{pr}} \approx {R_0}{e^{ - {t_n}/{\tau _0}}}$ is the success probability for distributing an entangled photon pair over the distance $L$.

We calculate the repeater rate for entanglement distributing between two remote quantum nodes. We assume $n = 4,{L_0} = 63km$, and the distance between two remote quantum nodes is about 1000\textit{km}. The total detection efficiency ${\eta _{TD}} \approx 90\% $ for the Stokes (anti-Stokes) detection channel. The memory lifetime ${\tau _0} = 16{\rm{ }}s$\cite{45} and quantum frequency conversion efficiency ${\eta _{FC}} = 46\% $\cite{46} are the best experimental performance at present as far as we know. With the retrieval efficiency at 80\% and the mode number at 100, the communication rate can reach 1Hz, which holds promise for long-distance quantum communication. 

\section{Demonstration of temporally multiplexed quantum repeater link}

\begin{figure}[htp]
	\centering\includegraphics[width=8.5cm]{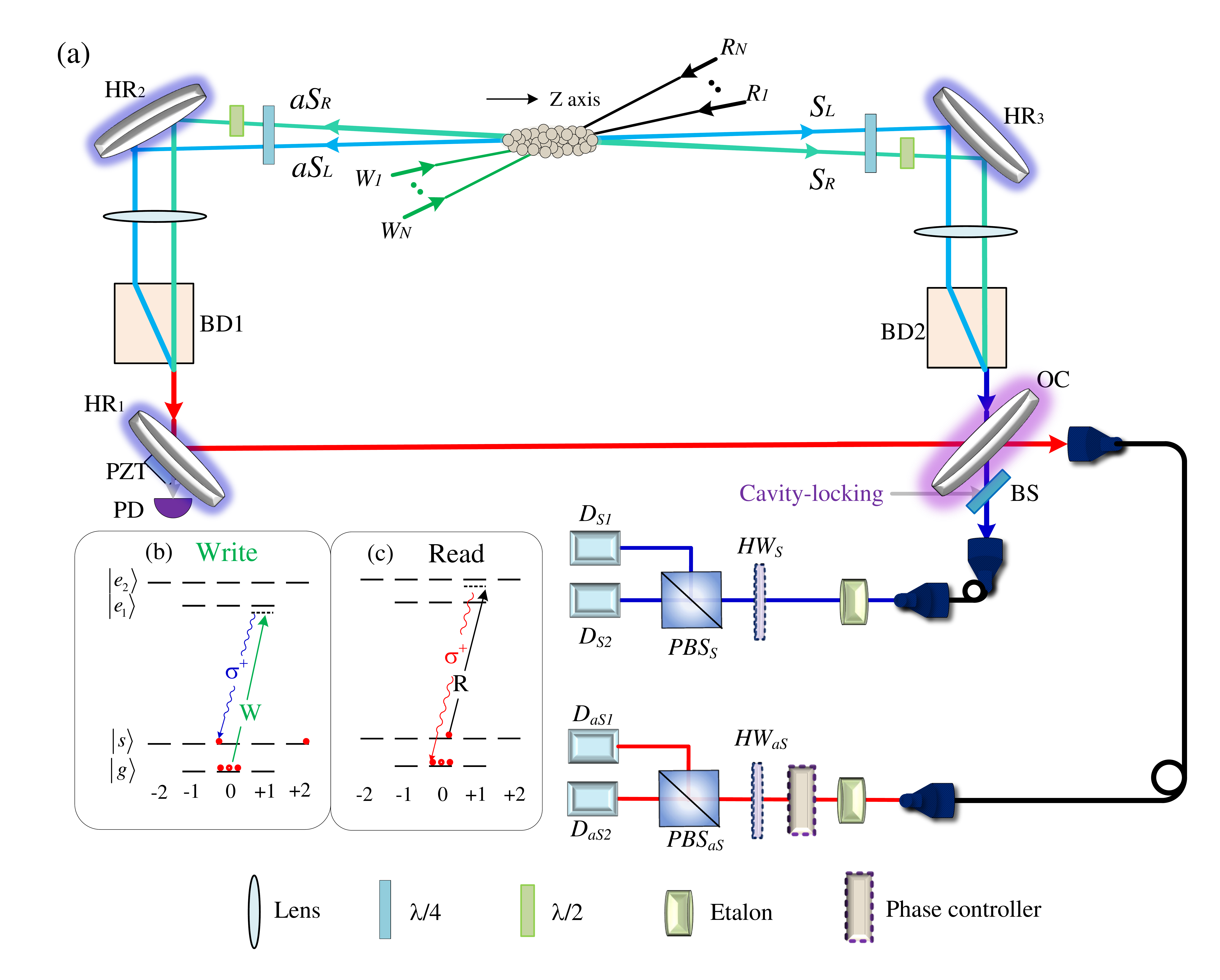}
	\caption{Schematic diagram of entanglement generation between two temporally multiplexed quantum memory nodes. (a) Experimental setup. A train of 12 write pulses with different directions is applied to the atoms to create temporally multiplexed quantum interface between atomic spin wave and Stokes photon via SRS. The Stokes fields S$_L$ and S$_R$ are collected at 0.14$^o$ to Z axis, which defines the spatial modes of the atomic ensembles L and R. The modes S$_L$ and S$_R$ are combined as the field S at the end of BD2. After the field S outputs from the cavity, it is collected by a single-mode fiber and sent to the measurement system. The entanglement between two atomic ensembles L and R is generated via single photon interference between S$_L$ and S$_R$ modes. To verity the entanglement between two atomic memories, the two spin waves are retrieved to aS$_L$ and aS$_R$ fields, respectively, by applying a corresponding read pulse. The fields aS$_L$ and aS$_R$ are combined as the field aS on BD1. Escaping from the cavity the field aS is sent to the measurement system. (b) and (c) are the relevant atomic levels involved in the write and read processes, respectively.}
\end{figure}

The memory medium is a $^{87}$Rb cold atomic cloud trapped in magneto-optical trap (MOT). A guiding magnetic field (B=4G) along Z axis is applied to define the quantization axis. A ring-cavity is placed around the atoms to improve the retrieval efficiency, and the details can be found in our previous work\cite{14}.

The atomic ground levels $\left| g \right\rangle  = \left| {5{S_{1/2}},F{\rm{ = 1}}} \right\rangle $ and $\left| s \right\rangle  = \left| {5{S_{1/2}},F{\rm{ = 2}}} \right\rangle $ together with the excited level $\left| {{e_{\rm{1}}}} \right\rangle  = \left| {5{P_{1/2}},F'{\rm{ = 1}}} \right\rangle $($\left| {{e_{\rm{2}}}} \right\rangle  = \left| {5{P_{1/2}},F'{\rm{ = 2}}} \right\rangle $) form a $\Lambda $ -type system, shown in Fig.2(b) and 2(c).The atoms are initially prepared in the Zeeman state $\left| {g,{m_{{F_g}}} = 0} \right\rangle $. A train of 12 write pulses is applied to the atoms to create the correlation between atomic spin wave and Stokes field via SRS. The Stokes photons excited by the \textit{i}-th write pulse will be detected in the \textit{i}-th measurement window. Each write pulse enters the atoms with different propagation directions. The angle between the two adjacent write pulses is 0.28$^o$, and the angle between the 1-st write pulse and Z axis is 0.2$^o$. The train lasts $\Delta T{\rm{ = }}8\mu {\rm{s}}$ and the interval between two adjacent write pulses is 400ns. All write pulses are ${\sigma ^{\rm{ + }}}$-polarized with red-detuned by 110 MHz to the $\left| g \right\rangle  \to \left| {{e_{\rm{1}}}} \right\rangle $ transition. Each write pulse induces the Raman transition $\left| {g,{m_{Fg}} = {\rm{0}}} \right\rangle  \to \left| {s,{m_{Fs}} = {\rm{0}}} \right\rangle $ via $\left| {{e_1},{m_{F{e_1}}} = {\rm{1}}} \right\rangle $, which emits a ${\sigma ^{\rm{ + }}}$-polarized Stokes photon and simultaneously creates an atomic spin-wave associated with the clock coherence $\left| {{m_{Fg}} = {\rm{0}}} \right\rangle  \leftrightarrow \left| {{m_{Fs}} = {\rm{0}}} \right\rangle $ probabilistically.

For constructing two atomic ensembles L and R with different spatial modes to simulate the two end nodes in an elementary link, the two Stokes fields S$_L$ and S$_R$ are collected at 0.14$^o$ to Z axis. After the interaction of the atoms with the \textit{i}-th write pulse, the joint state of the Stokes field and the associated spin-wave is described as:
\begin{equation}
{\left| {\psi _{}^i} \right\rangle _{L(R)}} = \left| {{{\rm{0}}_a}} \right\rangle \left| {{{\rm{0}}_s}} \right\rangle  + \sqrt \chi  {\left| {{\rm{1}}_a^i} \right\rangle _{L(R)}}{\left| {{\rm{1}}_s^i} \right\rangle _{L(R)}} + O(\chi )
\end{equation}
where, $\left| {{0_a}} \right\rangle$ ($\left| {{0_s}} \right\rangle $) denotes the vacuum part, $\chi $  is the excitation probability of one Stokes photon and it is the same for each write pulse in the experiment, ${\left| {{\rm{1}}_s^i} \right\rangle _{L(R)}}$  and ${\left| {1_a^i} \right\rangle _{L(R)}}$  are the Stokes field with one photon in mode S$_L$(S$_R$) and the single collective spin excitation in L(R) ensemble resulted by the \textit{i}-th write pulse. The single collective spin excitation can be written as
\begin{equation}
{\left| {{\rm{1}}_a^i} \right\rangle _{L(R)}}{\rm{ = }}\frac{1}{{\sqrt {{N_a}} }}\sum\limits_{{\rm{j}} = {\rm{1}}}^N {\exp ( - i\vec x_{j,L(R)}^ik_{a,L(R)}^i)} {\left| {{g_1}\cdots s_j^i\cdots {g_N}} \right\rangle _{L(R)}}
\end{equation}
where \textit{N$_a$} is the number of atoms, $\vec x_{j,L(R)}^i$  denotes the spatial position of the \textit{j}-th atom excited by the \textit{i}-th write pulse, $k_{a,L}^i = k_w^i - k_s^L$  ($k_{a,R}^i = k_w^i - k_s^R$ ) is the wave vector of atomic spin-wave, $k_w^i$  is the wave vector of the \textit{i}-th write pulse, and $k_s^L$ ($k_s^R$ ) is the wave vector of the Stokes photon in mode S$_L$(S$_R$).

In order to achieve entanglement between the two ensembles, we perform single photon interference on Stokes fields S$_L$ and S$_R$. The ${\sigma ^{\rm{ + }}}$ -polarized Stokes fields S$_L$ and S$_R$ are transformed into H- and V-polarization by a quarter wave-plate and a half wave plate, respectively, and then, they are combined as the field S on BD2. After the field S outputs from the cavity, it is collected by a single-mode fiber and sent to the measurement system, which includes a half-wave plate \textit{HW$_S$}, a polarization beam splitter \textit{PBS$_{S}$} and detectors \textit{D$_{S1}$} and \textit{D$_{S2}$}.At the two outputs of \textit{PBS$_{S}$}, the field S is changed into two modes $S_ + ^{} = \left( {S_R^{} + S_L^{}} \right)/\sqrt 2 $   and $S_{_ - }^{} = \left( {S_R^{} - S_L^{}} \right)/\sqrt 2 $  when  \textit{HW$_S$} is set at $\vartheta  = {22.5^o}$. The fields $S_ + ^{}$  and $S_ - ^{}$   are directed to detectors \textit{D$_{S1}$} and \textit{D$_{S2}$}, respectively. A click from \textit{D$_{S1}$} and \textit{D$_{S2}$} in the \textit{i}-th measurement window heralds that two ensembles are mapped into an entangled state: $\left| {\Psi _ \pm ^i} \right\rangle  = \left( {\left| {1_a^i} \right\rangle {}_R{{\left| 0 \right\rangle }_L} \pm {e^{ - i\phi }}\left| 0 \right\rangle {}_R{{\left| {1_a^i} \right\rangle }_L}} \right)/\sqrt 2 $ , where $\phi$ is the phase difference between two Stokes fields before they are combined on BD2.

To experimentally verify the entanglement state between the two ensembles, we map the atomic spin-wave into anti-Stokes fields by acting a corresponding ${\sigma ^{\rm{ + }}}$ -polarized read pulse. A feedforward system based on FPGA is utilized to determine the propagation direction of the read pulse. When a Stokes photon is detected in the \textit{i}-th measurement window, the read pulse with wave-vector $k_R^i =  - k_w^i$  is applied on the atoms by a feedforward controlled acousto-optic modulator (AOM). The atomic spin wave ${\left| {{\rm{1}}_a^i} \right\rangle _L}$  (${\left| {{\rm{1}}_a^i} \right\rangle _R}$) is transformed to ${\sigma ^{\rm{ + }}}$ -polarized anti-Stokes field aS$_L$(aS$_R$) with wave vector $k_{aS}^L =  - k_S^L$ ($k_{aS}^R =  - k_S^R$).The entanglement state between anti-Stokes photons in modes aS$_L$ and aS$_R$  can be written as $\left| {\psi _{aS}^ \pm } \right\rangle  = \left( {\left| {{1_{aS}}} \right\rangle {}_R{{\left| 0 \right\rangle }_L} \pm {e^{ - i(\phi  + \varphi )}}\left| 0 \right\rangle {}_R{{\left| {{1_{aS}}} \right\rangle }_L}} \right)/\sqrt 2 $ , where ${\left| {{\rm{1}}_{aS}^{}} \right\rangle _{L(R)}}$  is the anti-Stokes field with one photon in mode aS$_L$ (aS$_R$), and  $\varphi$  is the phase difference between two anti-Stokes fields before they are combined on BD1. By controlling the temperature of BD1 and BD2 precisely, the phase difference $\phi  + \varphi $  is kept constant.

Then, the fields aS$_L$ and aS$_R$ are transformed into H-polarized and V-polarized by wave plates and combined into the field aS on BD1. After the aS field escapes from the cavity, it is collected by a single-mode fiber and sent to the measurement system. The Stokes photon and anti-Stokes fields are resonated with the cavity. The cavity is locked via the Pound-Drever-Hall (PDH) method by using a locking laser pulse, which is coupled to the cavity mode through a beam splitter (BS, R=95\%).

\section{Experimental results}
For charactering the temporally multiplexed elementary link, we measure the concurrence of entanglement between two ensembles, which ranges from 0 for a separable state to 1 for a maximally entangled state. The concurrence is measured as $C = \max \left( {0,\frac{{2\left| d \right| - 2\sqrt {P_{00}^{}P_{11}^{}} }}{P}} \right)$  \cite{47}, $P_{mn}^{}$ corresponds to the probability to find m photons in the field aS$_R$  and n photons in the field aS$_L$  in the read process after the Stokes photons are detected, $P = P_{00}^{} + P_{01}^{} + P_{10}^{} + P_{11}^{}$,  $d = V\left( {P_{10}^{} + P_{01}^{}} \right)/2$ is the coherence term between the states $\left| {{1_{aS}}} \right\rangle {}_R{\left| 0 \right\rangle _L}$ and $\left| 0 \right\rangle {}_R{\left| {{1_{aS}}} \right\rangle _L}$, V is the visibility of the interference fringes between the anti-Stokes fields $aS_R$  and $aS_L$   when their relative phase $\theta $  is scanned.
\begin{figure}
	
	\centering
	
	\begin{minipage}[c]{0.25\textwidth}
		
		\centering
		
		\includegraphics[width=4.5cm]{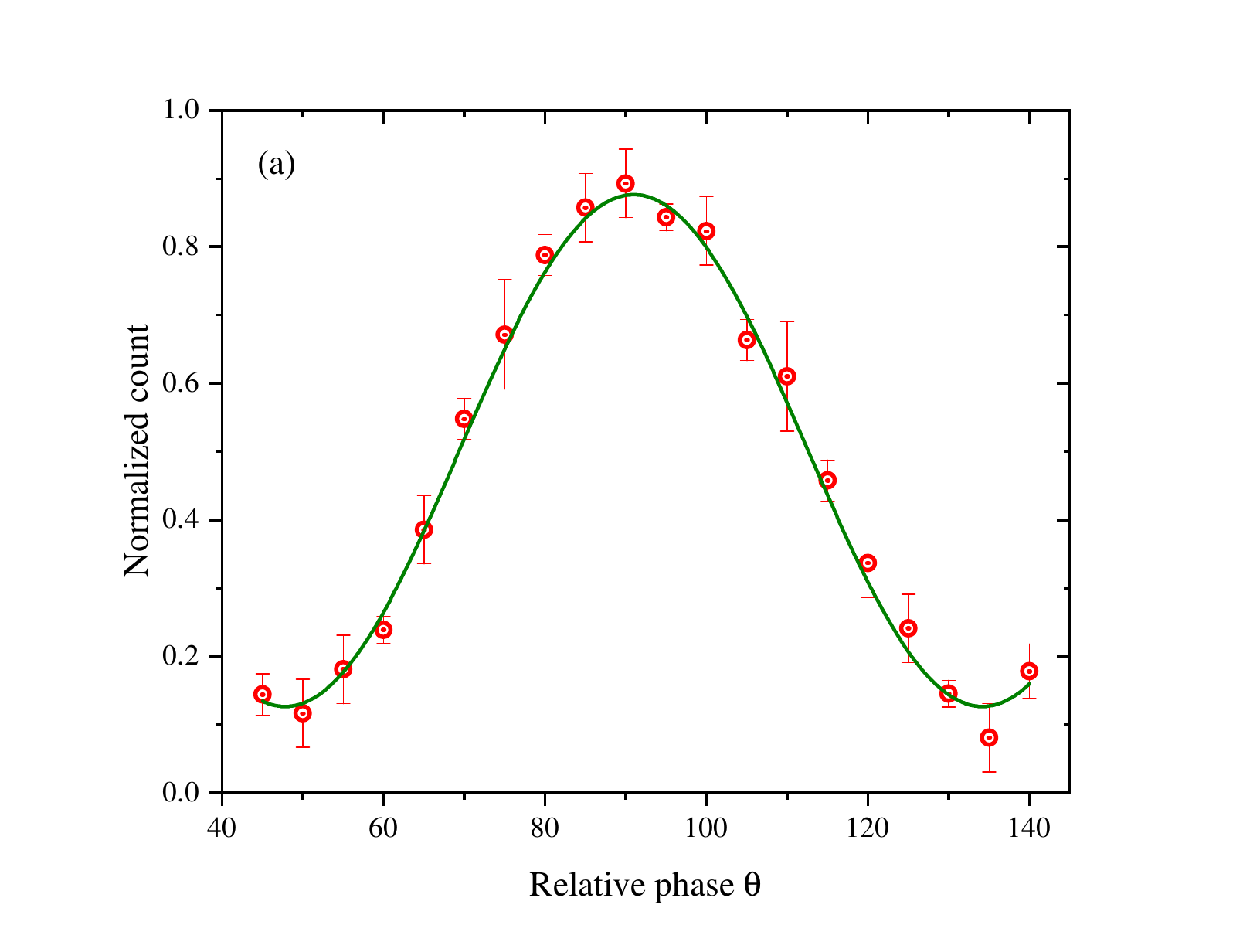}
		
	\end{minipage}%
	\begin{minipage}[c]{0.25\textwidth}
		
		\centering
		
		\includegraphics[width=4.5cm]{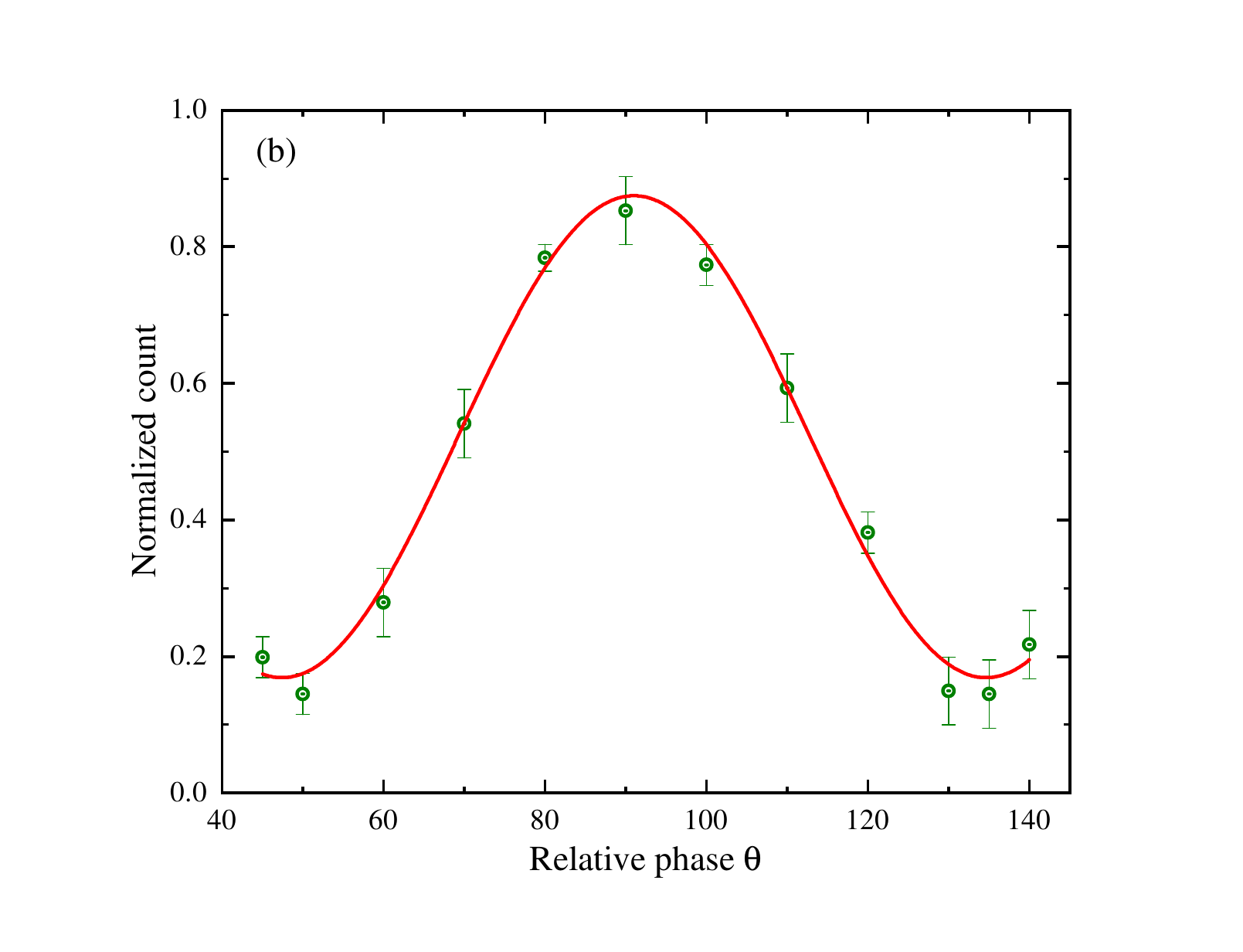}
		
	\end{minipage}
	
	\caption{The counts in detector DaS1 as a function of $\theta $ for storage time t=1 $\mu$s (a) and storage time t=150 $\mu$s (b).}
	
\end{figure}

To measure the interference visibility V, the polarization angle of \textit{HW$_{aS}$} is set at $\vartheta  = {22.5^o}$. The aS field evolves to $aS_ + ^{} = \left( {aS_R^{} + aS_L^{}} \right)/\sqrt 2 $  and $aS_ - ^{} = \left( {aS_R^{} - aS_L^{}} \right)/\sqrt 2 $   at the two outputs of\textit{PBS$_{aS}$}, which are directed into detectors \textit{D$_{aS1}$} and \textit{D$_{aS2}$}, respectively. We insert a phase controller before \textit{HW$_{aS}$}, which consists of two quarter wave plates and one half wave plate sandwiched in the middle. The relative phase $\theta $  between the anti-Stokes fields aS$_L$ and aS$_R$ can be changed by rotating the angle of the half wave plate. Along with the scan of $\theta $  , counts in detector \textit{D$_{aS1}$} vary as a sinusoidal function, as shown in Fig. 3. The visibility V can be measured as:$V = \frac{{{\rm{Max(}}{{\rm{C}}_{{D_{S1}},{D_{aS1}}}}{\rm{)}} - {\rm{M}}in{\rm{(}}{{\rm{C}}_{{D_{S1}},{D_{aS1}}}}{\rm{)}}}}{{{\rm{Max(}}{{\rm{C}}_{{D_{S1}},{D_{aS1}}}}{\rm{)}} + {\rm{M}}in{\rm{(}}{{\rm{C}}_{{D_{S1}},{D_{aS1}}}}{\rm{)}}}}$ , where, ${\rm{Max(}}{{\rm{C}}_{{D_{S1}},{D_{aS1}}}}{\rm{)}}$  (${\rm{Min(}}{{\rm{C}}_{{D_{S1}},{D_{aS1}}}}{\rm{)}}$) denotes the maximum (minimum) coincidence counts between the Stokes and anti-Stokes fields. Thus, we could get $V = 0.795 \pm 0.015$  at storage time t=1$\mu$s and $V = 0.7 \pm 0.024$  at storage time t=150 $\mu$s. For all experimental results the excitation probability $\chi $=1\%.  
\begin{table}[ht!]
	\centering\includegraphics[width=7.5cm]{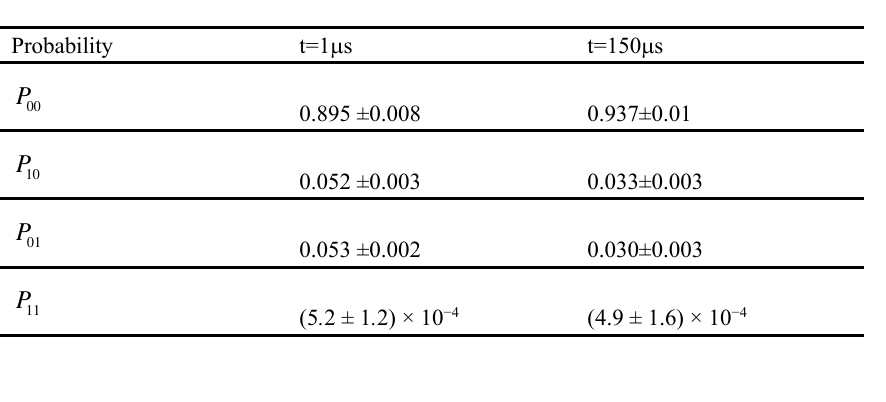}
	\caption{The measured $P_{mn}^{}$  at storage time t=1 $\mu$s and t=150 $\mu$s.}
\end{table}

To measure the  $P_{mn}^{}$, we set the polarization angle of \textit{HW$_{aS}$} is set at $\vartheta  = {0^o}$. The probabilities of $P_{mn}^{}$  can be directly measured via photon counting in the fields aS$_R$  and aS$_L$  using detectors \textit{D$_{aS1}$} and \textit{D$_{aS2}$}, as shown in table 1.

Based on the above measurements, the value of the concurrence of the entanglement between two ensembles can be deduced. The concurrence as a function of the storage time is shown in Fig.4. We get a concurrence of C=0.040(2) at storage time t=1 $\mu$s and C=0.001(2) at storage time t=150 $\mu$s for $\left| {\psi _{aS}^ + } \right\rangle $ . The experimental results show that the entanglement between the two ensembles can be remained for at least 150 $\mu$s.

\begin{figure}[ht]
	\centering\includegraphics[width=7cm]{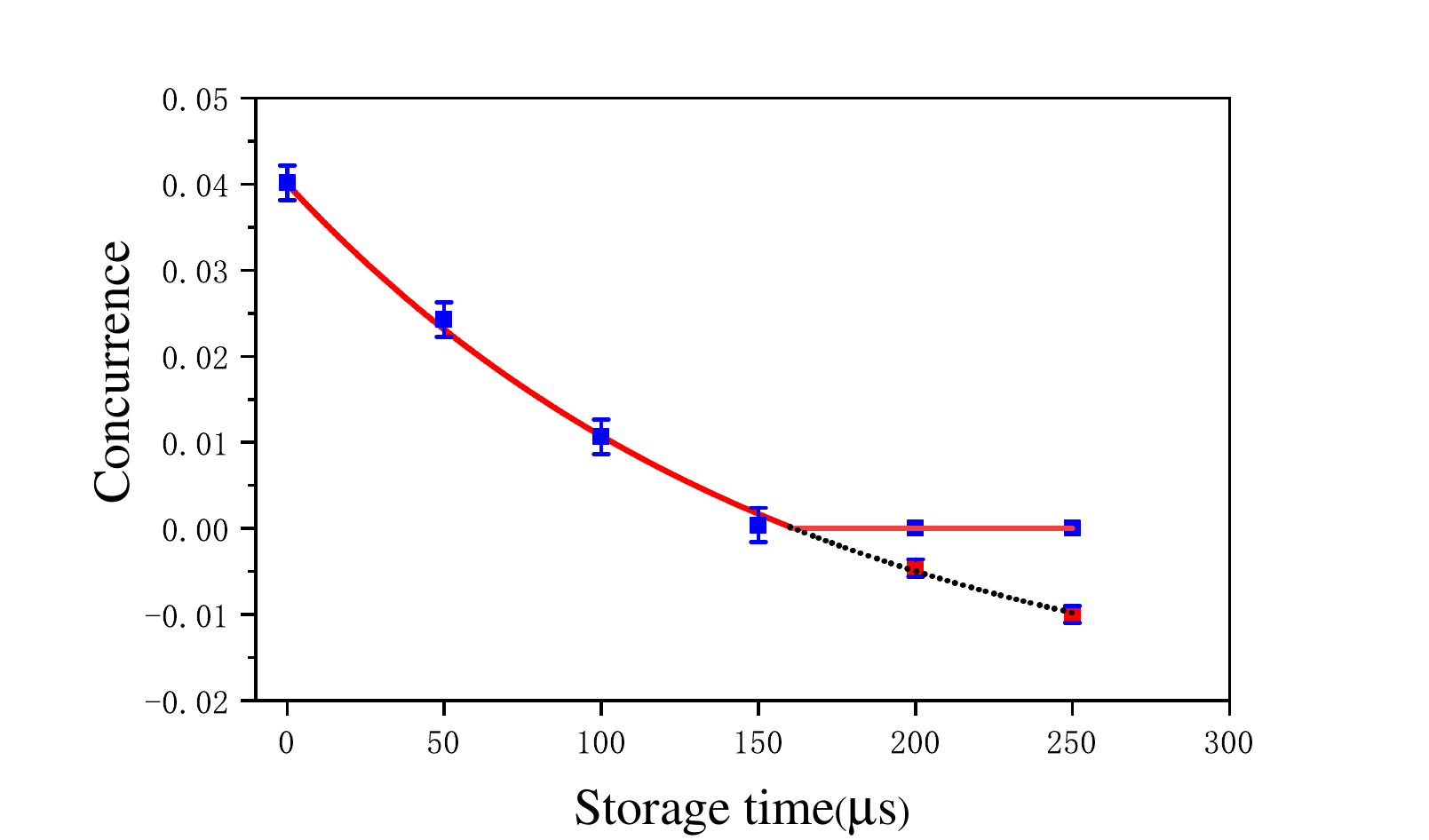}
	\caption{The measured concurrence C as a function of the storage time  for  $\chi $=1\%.}
\end{figure}

Meanwhile, we measure the intrinsic efficiency of temporal multiplexed quantum memory. The intrinsic efficiency defined by $\eta  = {{\left( {P_{01}^{} + P_{10}^{}} \right)} \mathord{\left/
		{\vphantom {{\left( {P_{01}^{} + P_{10}^{}} \right)} {\left[ {\sum\limits_i^{12} {\left( {P_{{D_{S1}}}^i + P_{{D_{S2}}}^i} \right)\;{\eta _D}\;} } \right]}}} \right.
		\kern-\nulldelimiterspace} {\left[ {\sum\limits_i^{12} {\left( {P_{{D_{S1}}}^i + P_{{D_{S2}}}^i} \right)\;{\eta _D}\;} } \right]}}$  , where, $P_{{D_{S1}}}^i$  ($P_{{D_{S2}}}^i$) denotes the probability of detecting Stokes photons at the detectors \textit{D$_{S1}$} (\textit{D$_{S2}$})  in \textit{i}-th measurement window, and ${\eta _D}$  is the total detection efficiency for anti-Stokes photons. Fig.5 plots the measured intrinsic efficiency (circles) as a function of storage time. We fit the data according to the function $R(t){\rm{ = }}{R_{\rm{0}}}\exp \left( {{{{\rm{ - t}}} \mathord{\left/
			{\vphantom {{{\rm{ - t}}} {{\tau _0}}}} \right.
			\kern-\nulldelimiterspace} {{\tau _0}}}} \right)$ (shown by the solid curve), which yields the intrinsic efficiency at the zero delay ${R_{\rm{0}}} = {\rm{70}}{\rm{.7}}\% {\rm{  }}$ , together with a memory lifetime ${\tau _{\rm{0}}} \approx {\rm{0}}{\rm{.3}}ms$ . 
 
\begin{figure}[ht]
	\centering\includegraphics[width=7cm]{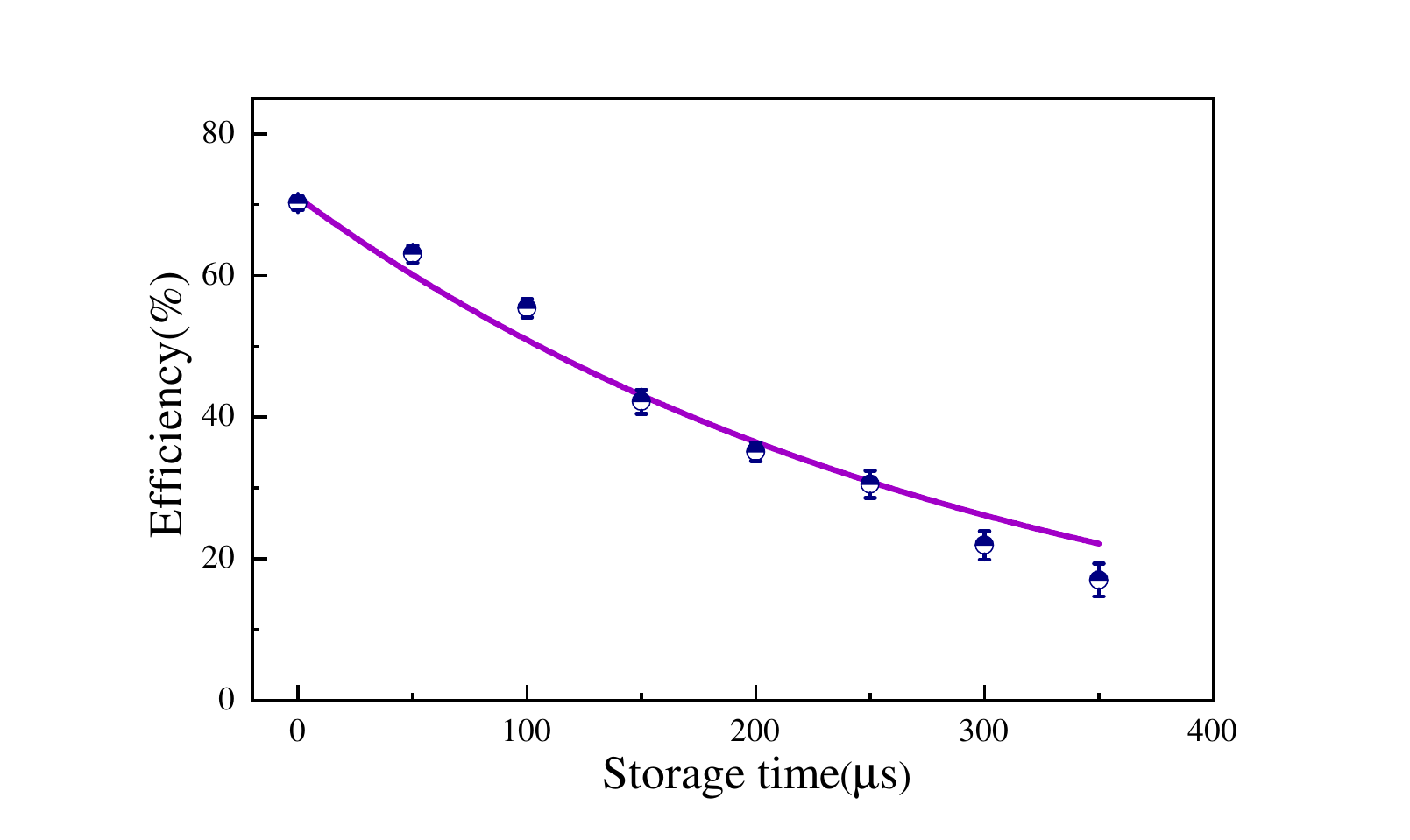}
	\caption{The measured intrinsic efficiency (circles) as a function of storage time . }
\end{figure}
Then, we present the entanglement generation rate (EGR), as a function of the multiplexed mode number N, shown in Fig.6. The EGR with n-mode multiplexing is proportional to the sum of the Stokes photon detection probabilities at the detectors  \textit{D$_{S1}$} and \textit{D$_{S2}$} $P_D^{(N)} = \sum\limits_i^N {\left( {P_{{D_{S1}}}^i + P_{{D_{S2}}}^i} \right)\;\;} $ . The result shows the EGR increases linearly with the number of modes, which is good agreement with the theoretical prediction $P_D^{(N)} = N{p_D}$ , where  ${p_D}$ is the EGR for single mode. when the number of modes is 12, ${{P_D^{(12)}} \mathord{\left/
		{\vphantom {{P_D^{(12)}} {p_D^{}}}} \right.
		\kern-\nulldelimiterspace} {p_D^{}}} = 11.79 \pm 0.35$ .

\begin{figure}[ht]
	\centering\includegraphics[width=7cm]{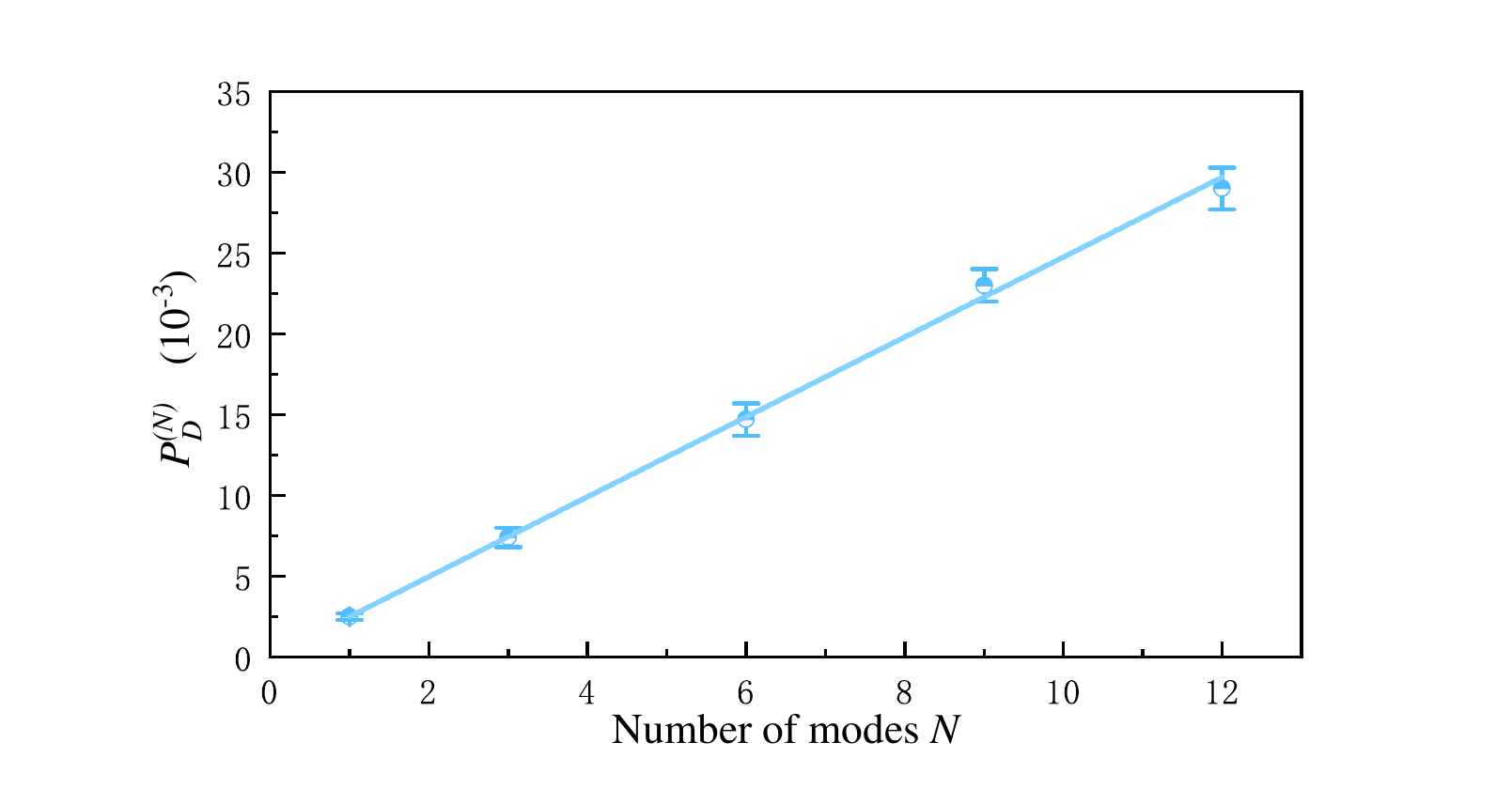}
	\caption{The detection probability of Stokes photon $P_D^{(N)}$  as a function of the number of modes at storage time t=1$\mu$s. The line is the linear fitting according to  $P_D^{(N)} = N{p_D}$ , where ${P_D} = 2.5 \times {10^{ - 3}}$  is the detection probability with single mode.  }
\end{figure}
\begin{figure}[ht]
\centering
\includegraphics[width=7cm]{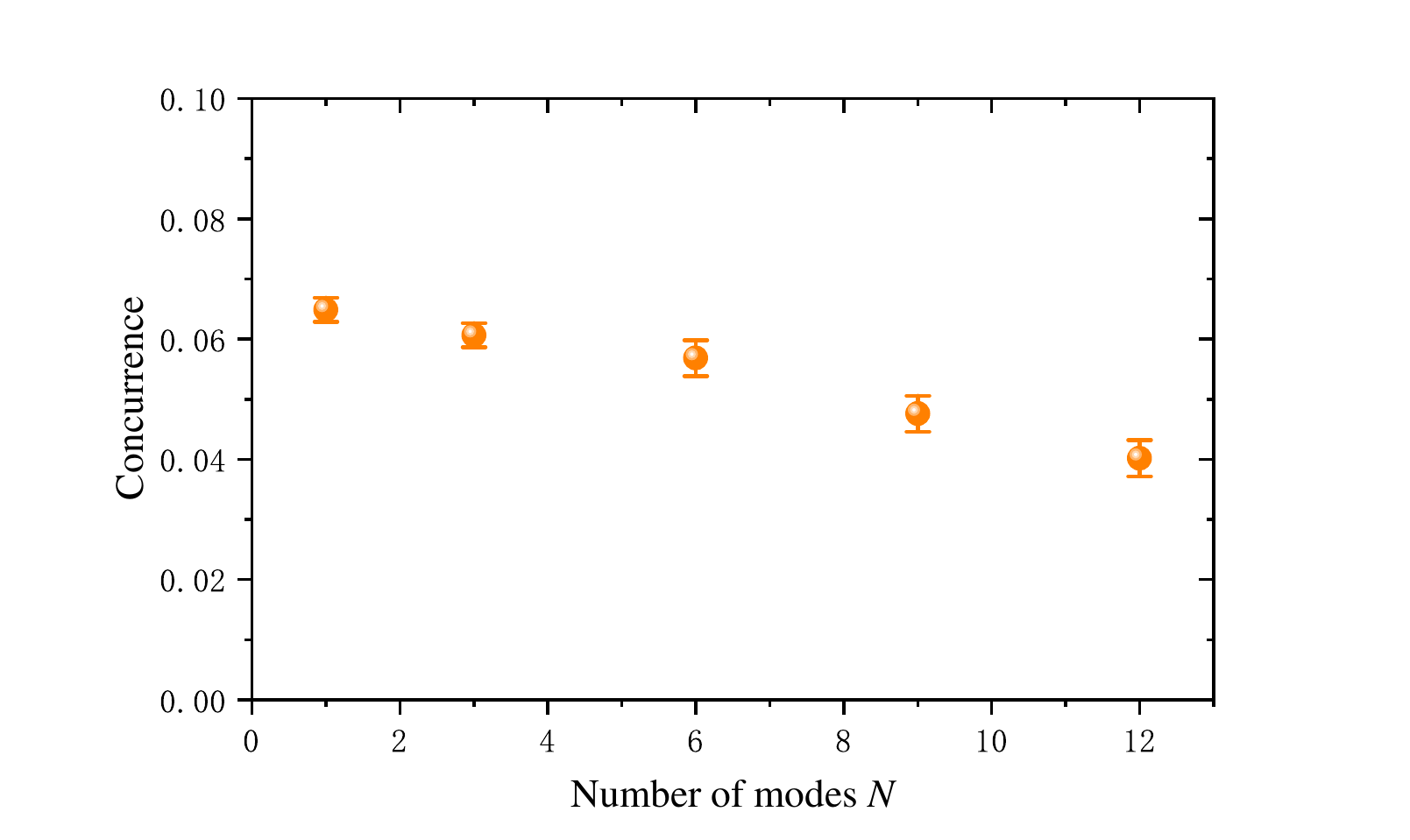}
\caption{The measured concurrence as a function of mode number at storage time t=1$\mu$s. }
\end{figure}
Finally, we also measure the concurrence as a function of number of modes at storage time t=1$\mu$s, which is showed in Fig.7. As the number of modes increases, the concurrence decreases. We attribute the decrease in concurrence to the crosstalk between the different modes. When one Stokes photon is detected in the \textit{i}-th measurement window, we retrieve the spin wave by the corresponding read pulse with wave vector $k_R^i$  in the read process. However, the phase-mismatched spin waves generated by all the other write pulses can also be converted photons with a certain probability. These non-directional emitted photons will enter the anti-Stokes fields  $aS_R$  and $aS_L$ , and cause a decrease in concurrence. In the future, We can suppress the background noise by using an asymmetrical photon-collection channel\cite{48} or designing the specially cavity which has different enhancement for Stokes photon and anti-Stokes\cite{34}.

\section{Conclusion and Discussion}
We demonstrate a temporally multiplexed elementary link based on atomic ensembles, wherein spin waves stored in atoms can be effectively transformed into photons on demand. We generate two atomic memories with distinct spatial modes by collecting emitted Stokes photons in two spatial modes, and then entanglement between the two quantum memories is created through single photon interference. Compared with single mode link, the rate of entanglement generation has increased by an order of magnitude (11.8-flod). By using the cavity-enhanced scheme, the on-demand retrieval efficiency of atomic spin waves is improved to 70\%, which is beneficial for the entanglement swapping between adjacent links.Combining frequency conversion\cite{46,49,50,51}, our multiplexing quantum repeater link with cavity-enhance holds promise for long-distance communication in optical fiber. To improve the entanglement rate further, we can load the atoms in optical lattice \cite{21, 45} to expand the storage lifetime. In addition, one can realize large-scale multiplexing capability by combining temporal and spatial multiplexing\cite{52}. The realization of temporal multiplexing quantum repeater link with high retrieval efficiency lays a foundation for the development of practical quantum networks.

\part*{{\normalsize {\Large Acknowledgements}}}

This work is supported by the National Natural Science Foundation of China (12174235), the Fund for Shanxi Key Subjects Construction (1331), and the Fundamental Research Program of Shanxi Province (202203021221011)

\end{document}